\documentclass[12pt,letterpaper]{article}
\usepackage[colorlinks,urlcolor=blue,citecolor=blue,linkcolor=blue]{hyperref}
\usepackage{natbib,graphicx}

\begin{document}
\pagestyle{myheadings}
\title{The ATA Commensal Observing System \\
\large{ATA Memo \#89}}
\author{Peter K. G. Williams
 (\href{mailto:pwilliams@astro.berkeley.edu}{pwilliams@astro.berkeley.edu})}
\date{2012 January 25}
\markright{\href{http://ral.berkeley.edu/ata/memos/}{ATA Memo \#89}
  --- ATA Commensal System}
\maketitle

\begin{abstract}
This memo describes the system used to conduct commensal correlator
and beamformer observations at the ATA. This system was deployed for
$\sim$2 years until the ATA hibernation in 2011 and was responsible
for collecting $>$5 TB of data during thousands of hours of
observations. The general system design is presented
(\S\ref{s:sdesign}) and the implementation is discussed in detail
(\S\ref{s:impl}). I emphasize the rationale for various design
decisions and attempt to document a few aspects of ATA operations that
might not be obvious to non-insiders. I close with some
recommendations (\S\ref{s:recos}) from my experience developing the
software infrastructure and managing the correlator
observations. These include: reuse existing systems; solve, don't
avoid, tensions between projects, and share infrastructure; plan to
make standalone observations to complement the commensal ones; and be
considerate of observatory staff when deploying new and unusual
observing modes. The structure of the software codebase is documented
(\S\ref{s:impl}, \S\ref{s:cmds}, \S\ref{s:modules}).
\end{abstract}

\section{Introduction}

One of the major design goals of the Allen Telescope Array
\citep[ATA;][]{theata} was the ability to share observatory resources
to conduct multiple independent observing programs simultaneously ---
a feat generally referred to as commensal observing.\footnote{In
  biology, commensalism is a symbiotic arrangement between mutualism
  and parasitism: it is the coexistence of two organisms in which one
  benefits and the other neither benefits nor suffers. It derives from
  the Latin {\it cum mensa}, ``sharing a table,'' originally alluding
  to the sharing of food scraps in particular.} This memo describes
the system used to conduct the largest commensal observing campaigns
before the ATA's hibernation in 2011. The primary campaign was a joint
survey of regions in the Galactic plane: a traditional
radio-astronomical survey looking for broadband Galactic radio
transients \citep[\textit{\`a la}][]{hlk+05} and a SETI (Search for
Extraterrestrial Intelligence) search for narrowband sources. In both
cases, the Galactic population is clearly best probed by looking in
the plane rather than searching the sky uniformly. Various components
of the traditional search have been referred to as the GCS (Galactic
Center Survey), the AGCTS (ATA Galactic Center Transient Survey), {\sf
  gal90} (Galactic plane, $l=90^\circ$ region), or the Kepler or
Cygnus X-3 surveys. These are now all grouped together as AGILITE, the
ATA Galactic Lightcurve and Transient Experiment. (The survey
description paper is currently in preparation.) Because of the low
overhead to doing so, commensal correlator observations were also made
during the SETI ``exoplanets'' and ``waterhole'' surveys, but there
are currently no specific plans to use the correlator data from these
undertakings.

The original vision that motivated the goal of substantial commensal
observing on the ATA was one in which scheduling was based on
traditional radio astronomy applications but SETI searches ran
continuously in the background as well. The traditional observing
program would drive the telescope, controlling the pointing and most
hardware systems, and use the ATA's correlator backends to take
data. Meanwhile SETI observations would be performed using the ATA's
beamformer backends more-or-less passively, choosing any promising
observation targets lying within whatever field of view (FOV) was
chosen by the traditional observing program. It's worth noting that
this is not the only commensal observing scheme that might feasibly be
implemented at the ATA. For instance, with a sufficiently large number
of dishes and two relatively undemanding traditional radio
astronomical observing projects, one could partition the array into
two subarrays and run the two projects simultaneously, each using
separate antennas and correlators. (One could argue that this is in
fact not a commensal mode since so few resources would be shared
between the two projects.)

The earliest commensal observations at the ATA were performed in
August 2008 by Karto Keating and Tom Kilsdonk, but these and a few
other efforts never became routine observing modes. As such, I refer
to the system under discussion as ``the ATA commensal observing
system,'' without further qualification. A design for an
observatory-wide commensal observing system, and a deeper discussion
of the observatory-wide software systems, is presented in
\citet{gkkm+10}.

\section{Survey and System Design}
\label{s:sdesign}

Observations for the commensal campaigns are scheduled in blocks as
per the standard system in use at the ATA. In contrast to the
long-term vision of ATA commensal observations, it is SETI, not the
traditional radio observing program, that's ``in the driver's seat''
for the observations: SETI software takes responsibility for all of
the telescope hardware, most importantly the pointing. This
arrangement came about because SETI already had a well-established
observing system called Prelude, which had been adapted from its
Arecibo roots to work at the ATA as well. Given this existing
codebase, the project was approached with the plan of minimizing the
amount of changes required to Prelude, while adding a separate
``commensal observer'' component that would take care of everything
related to the commensal correlator campaigns.

With Prelude in charge of pointing the antennas and taking care of
SETI's data-taking, the responsibilities of the commensal component of
the observing campaigns are {\it extremely} constrained: essentially,
all it can do, and all it needs to do, is turn the correlators on and
off. One could envision a much more complex system in which the
commensal observer dynamically notifies Prelude of various needs
(``please visit a phase calibrator''), but the system is vastly
simplified if all such decision-making is centralized in Prelude. This
simplification was made possible by pre-negotiating such decisions as
pointing centers, dwell times, calibrator intervals, and focus
settings.

In principle, the commensal observer could be completely ignorant of
Prelude and its workings; by monitoring the current pointing
directions of the ATA dishes, it could decide when a source was being
observed and when the array was slewing, and in the former case it
could do the necessary work to take data. It turns out, however, that
``where is the ATA pointing?'' and ``are we tracking or slewing?'' are
questions that are more difficult to answer than one might think:
obtaining ATA antenna telemetry requires fairly complex code (cf. the
implementation of the {\sf atastatus} command), and glitches in the
data stream make it difficult to interpret the data robustly. Prelude
knows exactly what its intentions are, however, so it was decided that
the commensal observer would monitor a telemetry stream from Prelude
to obtain pointing information. This stream is described more fully
below (\S\ref{s:synch}).

An important thread running through the design of the commensal system
is automation: with thousands of hours of commensal observations
scheduled, it's desirable to execute them using as little human
intervention as possible. The existing Prelude system fortunately
dealt with the difficult task of choosing targets and planning a
night's observing.

\section{System Implementation}
\label{s:impl}

The ATA commensal observing system is composed of a group of software
tools and practices for using them. As mentioned above, some aspects
of the commensal system were implemented in Prelude and its successor,
SonATA (``SETI on the ATA''). This code is internal to SETI and is not
discussed here.

Although the rest of the commensal software runs on a diverse set of
hosts in a diverse range of situations ({\it e.g.}, at HCRO during
observing; at UCB during data reduction), the whole codebase is
maintained in a single Git (\url{http://git-scm.com/}) repository by
the author. This document describes Git commit {\sf
  88263be4\-4c3d724e\-700e46b1\-56e9b1dc\-fd0b1089}, made on 2011 November
17. (Due to the nature of the Git version control system, this one
value identifies an exact snapshot of the commensal source tree as
well as its entire revision history.) A public version of the
commensal repository is currently available at
\url{http://astro.berkeley.edu/~pkwill/repos/commensal.git}. This URL
will likely go stale as I am (hopefully\ldots) soon leaving
Berkeley. A public copy of the repository may be established at his
GitHub account, \url{https://github.com/pkgw/}, and at a minimum the
repository will be available upon request. (Once again, thanks to the
design of Git, each repository stands alone and contains every version
of every file as well as the complete revision history of the source
tree.)

\begin{figure}
\includegraphics{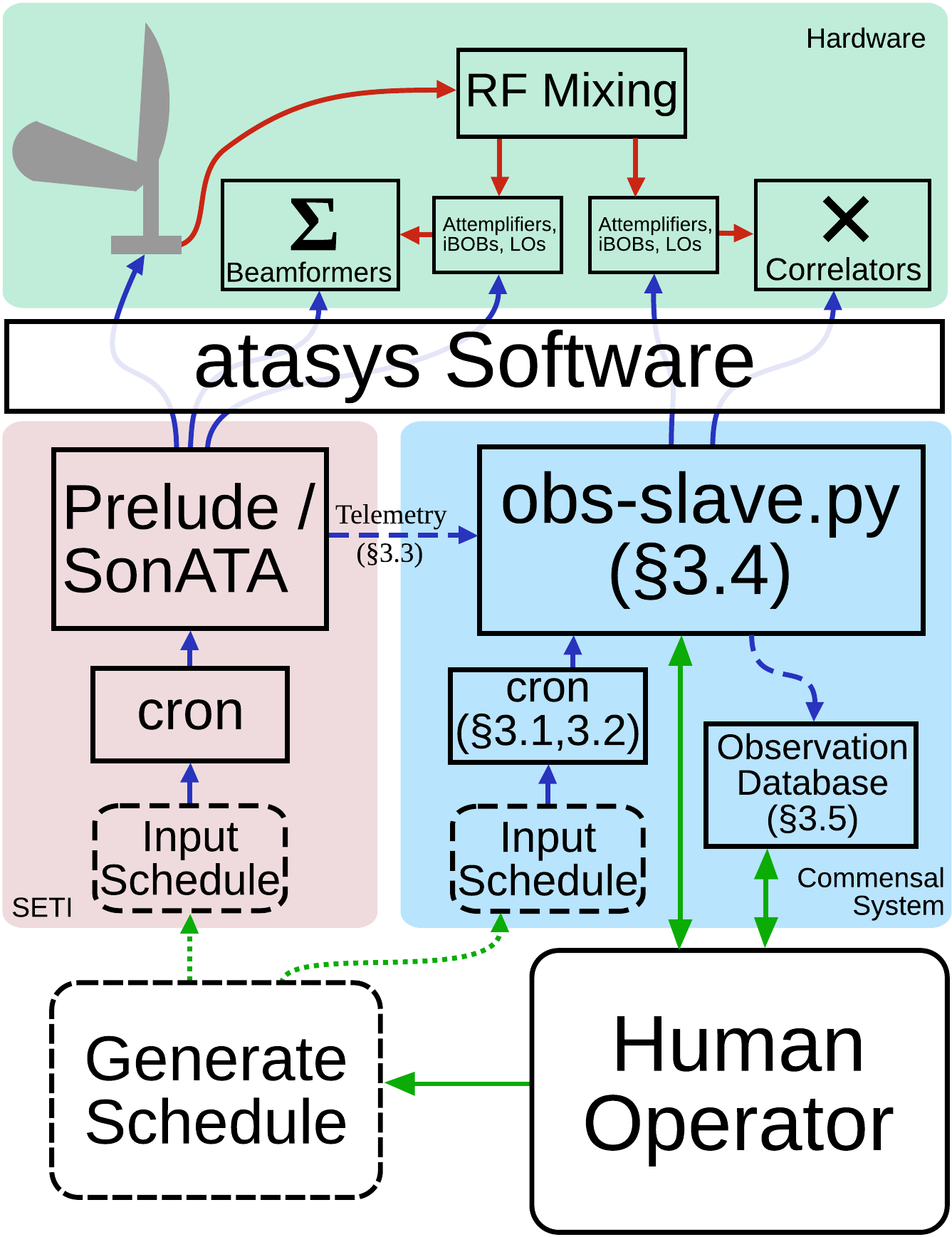}
\caption{Schematic of the commensal observing system. See
  \S\ref{s:impl}.}
\label{f:diagram}
\end{figure}

The vast majority of the source code is written in Python, with some
tools written in shell script. The majority of the latter are Bourne
shell scripts (most likely only compatible with {\sf bash}), but a few
are {\sf tcsh} scripts, since the latter was the language used for
most ATA observatory infrastructure. The repository also contains
scheduling and analysis metadata. The Python scripts reference some
support modules for controlling the ATA that are distributed in the
{\sf mmm}\footnote{The name derives from our weekly data analysis
  meeting, ``MIRIAD Masterclass with Melvyn'' (Wright).} repository at
\url{https://svn.hcro.org/mmm/pwilliams/pyata/}. A few secondary
packages use NumPy, miriad-python
(\url{http://purl.oclc.org/net/pkgwpub/miriad-python}\footnote{Note
  the unusual host name in this URL. This link is a ``Permanent URL'',
  one that is intended to be stable over decade timescales and thus
  (hopefully) more appropriate for inclusion in the academic
  literature. This is accomplished merely by forwarding requests for
  the PURL to a changeable destination, somewhat like the URL
  shorteners that are currently popular. I'm not using PURLs for most
  of the links in this paper, but this one in particular already
  existed because of its publication in \citet{wb10}. See
  \url{http://purl.org/} for more information.}) and/or my plotting
package, Omegaplot (\url{https://github.com/pkgw/omegaplot/}). The
system design is shown in schematic form in Figure~\ref{f:diagram}.

\subsection{Scheduling}

As mentioned above, commensal observations are allocated blocks in
the ATA schedule as per standard practice.

At the time of the observations, the observatory-wide system for
enacting this schedule was fairly {\it ad hoc}. The main HCRO
computing environment is a distributed Unix-type system, with user
logins and home directories shared over the network to a variety of
workstations and servers in the signal processing room (SPR). One user
account, {\sf obs}, has a shared password and by convention is used to
conduct the vast majority of observatory operations. SETI software
systems, however, are segregated from this shared system, and
generally run on separate machines with distinct logins. I'm not
familiar with their configuration.

Correlator observations are scheduled by constructing a large shell
script in the {\sf obs} account, {\sf obs.run}, that serially invokes
other shell scripts to perform standard observing tasks. Before the
2011 hibernation, most observatory time was devoted to correlator
observations so this could be considered the standard ATA scheduling
system. For purely SETI observations, the {\sf obs.run} script idles
for a fixed amount of time while the array is controlled from SETI
computers.

Because the {\sf obs.run} system proved to be unreliable and the
commensal campaigns involved many identical observations, the
commensal observer was designed to be scheduled and launched
separately in a more automatic fashion. Commensal observing blocks are
entered into a separate schedule file in the repository ({\sf
  sched/current.tab}) using the commands {\sf plan-schedsess} or,
later, {\sf plan-importtxt}. The observing processes are launched via
{\sf cron} (see below).

Meanwhile, the SETI observations are launched via a {\sf cron} system
as well in the SETI part of the network. This means that there are in
effect at least {\it four} relevant versions of the ATA schedule: the
textual/graphical version distributed to users, that schedule as
incarnated in {\sf obs.run}, the schedule used by the commensal
observer, and the schedule used by Prelude. The large PiGSS (Pi GHz
Sky Survey) transient survey also used its own scheduling system. The
suboptimality of this situation should be obvious. Regarding the
particular case of the commensal campaigns, however, things weren't
too bad. While it was sometimes necessary to coordinate schedule
changes with SETI staff over email, and mistakes could result in lost
observing opportunities, most observations went smoothly. Rarely,
miscommunications would result in {\sf obs.run} and the commensal
launcher attempting to observe simultaneously; due to poor lockouts in
the observatory infrastructure, it's actually possible for both
systems to operate at the same time, with expectedly poor results.

Each observing session scheduled in the commensal system was assigned
a random 32-bit hexadecimal identifier, {\it e.g.} {\sf
  8592a0ce}. This identifier was used for tracking data and metadata
throughout the rest of the observing and analysis chain. (The original
scheme, used in the 2009 GCS survey, used universally unique
identifiers [UUIDs], which are a particular kind of 128-bit
identifier. They are 36 characters long and were judged to be
overkill.) Note that identifying a session by the day on which it
occurs, for instance, is insufficient if there are multiple sessions
in one day. In theory, it doesn't even work to identify a session by
its start time, since the array could be partitioned into two
subarrays with different sessions being executed during the same block
of time.

\subsection{Launching and Managing Observations}

The main challenges in automatically launching commensal observations
are robustness and control. The loose nature of the ATA online system
makes it possible for the commensal observations to start up with the
array in odd states, and there are inevitably network outages,
computer crashes, hardware problems, {\it etc}. Meanwhile, the array
schedule isn't written in stone, and the operator needs to be able to
see what's going on and potentially alter array operations at will.

Commensal correlator observations are launched automatically under the
author's user account, {\sf pkwill}. A launcher program, {\sf
  obs-hourly}, runs hourly via {\sf cron} (see {\sf
  resources/obs.crontab}) on the host {\sf obscntl} and consults the
schedule file in a checkout of the repository found in {\sf pkwill}'s
home directory. If an observation is upcoming in the next hour, the
launcher stays resident and {\sf exec}s the appropriate observing
script at the right time. Because most observations are scheduled for
at least a few hours, the hourly invocation of the launcher was
intended to provide some insurance if the observing script crashes;
this feature has been helpful only a handful of times.

Observations can also be ``kickstarted'' manually if the automatic
invocation fails. This usually happens when an unusual hardware
condition needs to be resolved before correlator observations can
start. This is done with the {\sf obs-kickstart} command, which uses
GNU {\sf screen} to launch the observing script in a detached,
persistent virtual terminal. Both launchers have interlocks to attempt
to prevent multiple observations from running simultaneously and to
prevent observations from being launched outside of a scheduled
commensal observing block.

Due to the distributed nature of the HCRO computing environment,
observations can in theory be run from a variety of hosts. This
capability was not usually exploited in ATA operations. Commensal
observations were designed to be run from the {\sf cron} host {\sf
  obscntl} (AKA {\sf auxcntl}) or, in early days, {\sf tumulus}.

The {\sf obs} user can monitor and control commensal observations with
the command {\sf commensal-mc}, which was set up to be easily
accessible to the {\sf obs} user and was intended to be
straightforward to use. The command can perform diagnostics such as
checking whether a commensal observing session is currently active
(according to the commensal schedule), whether any commensal observing
processes are currently active, killing or kickstarting observer
processes, {\it etc}. Communication between processes and users is
accomplished with special files in the shared ATA archive disk, {\sf
  /ataarchive}, which is accessible to all HCRO computers and is
writeable by most user accounts (notably, {\sf obs} and {\sf
  pkwill}). Commensal observing processes should be run as the user
{\sf pkwill}, which presents a difficulty if the user {\sf obs} is
attempting to kickstart a new process. This is dealt with by doing the
kickstarting through a special passwordless SSH key with a preset
command.

A more detailed look at the data coming out of a commensal observation
can be obtained with {\sf misc-parseplog}, which parses log output
from the commensal observer and does some simple diagnostics on the
datasets being written by the data catchers. This program has proven
to be quite helpful for checking target selection and catching
occasional correlator failures which were both serious and undetected
by the usual diagnostics.

\subsection{Synchronization Protocol}
\label{s:synch}

As alluded to above, the commensal observing software monitors a
telemetry stream from Prelude. This stream was designed to convey the
minimal amount of information necessary for the commensal observer to
easily drive the correlators. The stream is broken into messages, each
of which transmits a timestamp, an array state, and observing
coordinates. The states are listed in Table~\ref{t:syncstates}.

\begin{table}
\begin{tabular}{llp{4in}}
Name & Code & Description \cr
\hline
{\sf idle} & 0 & Observing is not active, and the coordinates are
  undefined \cr
{\sf slewing} & 1 & The telescopes are slewing, and the coordinates
  specify their destination \cr
{\sf pointed} & 2 & The telescopes are pointed up on the specified
  coordinates
\end{tabular}
\caption{Array states enumerated in the synchronization protocol.}
\label{t:syncstates}
\end{table}

This architecture lends itself to a state-machine implementation in
the commensal observer. The lack of state in the telemetry stream
means that the observer and/or Prelude can die at arbitrary times and
the observer will recover sensibly.

The telemetry stream is implemented with broadcast UDP packets on port
24243. (This port was chosen arbitrarily.) Each packet consists of 32
bytes: a 64-bit double-precision floating point Unix timestamp, a
32-bit unsigned integer state code, an unused 32-bit padding field, a
double-precision right ascension coordinate measured in hours, and a
double-precision declination coordinate measured in degrees. The
values are big-endian and can be decoded with the Python {\sf
  struct.unpack} specifier {\tt >dIIdd}. The numerical codes
corresponding to the states are also listed in
Table~\ref{t:syncstates}. The specified transmission rate is one
packet per second, though there's nothing special setting this
rate. Protocol decoding is implemented in {\sf pylib/protocol.py}.

Clearly, it is not difficult to implement this protocol. The program
{\sf obs-listener.py} can monitor the telemetry stream and print out
low-level packet contents and check for unexpected state
transitions. For example, it was used to diagnose a situation in which
two Prelude processes were simultaneously running, one broadcasting a
series of idle packets, the other attempting to run observations. The
apparent oscillation between the idle and active states
unsurprisingly confused the observer.

The program {\sf obs-master.py} can be used to perform commensal-style
observations when Prelude is inactive. Like Prelude, it performs
target selection, drives the telescopes, and emits telemetry
packets. Thanks to the clear division of responsibilities and the
simplicity of the commensal observing protocol, the commensal observer
can operate completely normally even though the software ``in the
driver's seat'' is completely different. This was also the experience
when Prelude was replaced with SonATA, SETI's next-generation
observing system. Besides a change in the active port number (to allow
simultaneous testing of Prelude and SonATA), the replacement of several
racks of computer equipment and the entire software codebase was
completely transparent to the commensal observer.

Compared to most elements of the commensal software package, the {\sf
  obs-master.py} program is relatively complicated. It can either loop
through a list of targets in order, with calibrator scans
interspersed, or it can always observe the lowest-declination visible
source. A few other parameters, such as the antenna focus position and
interval between calibrator scans, are also configurable. It was not
used extensively so its target-selection algorithms are relatively
unsophisticated.

The current code to monitor broadcast UDP packets doesn't work in the
case that the emitter and listener are on the same host, which
corresponds to situations when {\sf obs-master.py} is being used. I
suspect that this could be fixed with a better understanding of the
proper low-level socket setup, but since this problem arises during
manually-arranged observations, during those times I just tweak the
socket code to skip actual network communication (by using the
loopback device). This ``solution'' did lead to problems once or twice
when I forgot to undo the tweak once observations were over.

\subsection{Correlator Observer}

The commensal correlator observer, {\sf obs-slave.py}, is the most
complex element of the commensal observing system. It is a single
program that is launched by the hourly cronjob, monitors the telemetry
stream, and operates the ATA correlators to take data for transient
searches.

On a low level, the commensal observer is responsible for driving a
specific set of subsystems:
\begin{itemize}
\item the ATA correlators
\item the attemplifiers used by those correlators
\item the local oscillators (LOs) used by those correlators
\item the fringe-rotator iBOBs
\end{itemize}
The observer must take appropriate action based on state transitions
in the telemetry stream or other external inputs ({\it e.g.}, abort
commands).

The observer must also pay attention to some secondary systems to
support its observations:
\begin{itemize}
\item It must have access to a catalog (mapping from coordinates to
  source names) so that correct target names can be shown to the
  operator and embedded in output datasets.
\item It must generate ATA ``ephemeris files'' to be used by the
  fringe rotators and correlator data-catchers.
\item It must check for abort signals, which is accomplished by
  monitoring the {\sf /ataarchive} filesystem for the appearance of
  special files.
\item It must generate detailed logs, since one wants to be able to
  debug subtle problems, check that hardware is being driven
  correctly, search for efficiency losses, and monitor the array.
\item It must save appropriate state so that if it crashes and is
  restarted, the array hardware is reset {\it or not reset} so that
  datasets from before and after the crash may be calibrated
  consistently.
\end{itemize}

In order to accomplish all this, the observer is a multithreaded
program with a main coordinator thread and various subthreads
responsible for the subsystems. Because it is important that the
observer be resilient to crashes, there's also complex code to deal
with Python exceptions as robustly as possibly. The final program is
still less than 700 SLOC (statement lines of code), a nice
demonstration of the concision of Python. Looking at the source will
confirm, however, that the code is quite dense.

The commensal observer stores each individual scan in a separate
MIRIAD dataset simply named with a serial number. This approach makes
a directory listing relatively opaque, but was hoped to be more robust
than the standard ATA practice of naming datasets with some
combination of source name, observing frequency, {\it etc.}, and
appending to these datasets when multiple scans occur. Most MIRIAD
tasks can stream through multiple datasets, but none can truly analyze
only a portion of a dataset, so it should be more efficient to create
many datasets and sometimes deal with them in bulk, rather than to
create large datasets and sometimes subdivide them. There have also
been instances where a large dataset has been rendered invalid due to
some problem at the very end of an observation, and subdividing
datasets helps minimize the damage incurred in these cases.

The datasets generated by the commensal observer are augmented with a
few extra data items. A file within the dataset named {\sf c\_instr}
records the identifier of the ATA correlator used to generate the
data; there is perhaps an oversight in the ATA data catcher software
that this information is not recorded in datasets otherwise. (This
information is useful because failures in the digital electronics have
correlator-specific effects.) Another file named {\sf c\_tstop}
records the timestamp at which Prelude reported leaving the current
pointing for another target --- there will be a delay between Prelude
issuing the slew command and the correlator data catchers shutting
down and closing their datasets, so there may be some bad data at the
end of a scan taken while the antennas are slewing.

One particular challenge faced by the commensal observer is that the
standard ATA data catcher program, {\sf atafx}, was designed to be run
with a predefined integration time. By the nature of the commensal
observing system, however, the commensal observer does not know how
long each integration will last. (If the expected integration time
were transmitted from Prelude to the observer, one would still have to
check that it was honored, and dealing with unexpected circumstances
would require all of the flexibility needed by the current
implementation.) Given the current software, the best solution is
actually to SSH to {\sf strato}, the data-taking host, and kill the
{\sf atafx} processes. A project was started to change the data takers
to be always-running servers so that integration could be stopped and
started quickly and on-demand, but that code never reached deployment.

\subsection{Observation Database}

The commensal observing system includes not only software for making
observations but also a set of tools for understanding the
observations that have been made. These are built on a database of
observing sessions and individual scans.

These post-observation tools could plausibly have been separated into
a different software package, and that might arguably have been a
better design decision. Different observing programs may have different
post-observation analysis needs and thus could benefit from multiple
post-observation toolsets. On the other hand, any post-observation
analysis toolset requires some knowledge of available datasets and the
context in which they were obtained, so there's an advantage to
grouping this code with the observing code. It seemed helpful to not
splinter the relevant software into too many pieces, so the two
components were kept together.

That being said, an important aspect of the design of the
post-observation analysis toolkit was a reliance on only the datasets
stored in the ATA archive and no additional metadata. The reasoning
was that while one might have plenty of expectations about the data on
disk from the observing plan or even observing logs, the data on disk
are the ``ground truth,'' and there are always unexpected ways for the
logged metadata and recorded visibilities to disagree. There were
indeed many cases in practice in which the metadata and the actual
datasets disagreed.

For each correlator observing campaign, a table of observing sessions
and individual scans is maintained. The session table is populated
from the schedule with the program {\sf pdb-stubsess}, with new
sessions marked as ``unscanned''. After being observed, each session
is eventually scanned with {\sf pdb-scansess} and marked as
``scanned''. The {\sf pdb-scansess} program creates entries for each
scan in the session and in fact reads all of the visibility files
completely to search for any potential problems in the data. It also
records useful per-scan metadata, the fields of which are listed in
Table~\ref{t:scandata}. Of particular note are the {\sf lst0} field,
which allows quick analysis of the hour-angle coverage of observations
of a source, and the {\sf failtype} field, which records any issues
that make the scan unusable. To paraphrase {\it Anna Karenina},
``Successful observations are all alike; every unsuccessful
observation is unsuccessful in its own way.'' Thus there's one value
of {\sf failtype}, zero, which indicates success, but a variety of
nonzero values indicate possible failure modes, as in the familiar
{\sf errno} values returned in Unix error codes.

\begin{table}
\begin{tabular}{lllp{3in}}
Name & Type & Units & Description \cr
\hline
{\sf uid} & string & & UID of the scan's session \cr
{\sf fncode} & int & & identifies the scan filename within the session
  data directory via reference to a separate table \cr
{\sf ccode} & int & & identifies the equatorial coordinates of the scan
  pointing via reference to a separate table \cr
{\sf freq} & int & MHz & the observing frequency of the scan \cr
{\sf focus} & int & MHz & the predominant focus setting of the
  antennas during the scan \cr
{\sf tst} & int & seconds & the start time of the scan (as a Unix
  time-since-Epoch) \cr
{\sf dur} & int & seconds & the duration of the scan \cr
{\sf lst0} & float & hours & the LST at the beginning of the scan \cr
{\sf failtype} & int & & failure flag if scan is unusable for
  some reason
\end{tabular}
\caption{Metadata recorded in the scan database.}
\label{t:scandata}
\end{table}

The observation databases are recorded in flat, line-oriented text
files using a simple database layer implemented in {\sf
  pylib/flatdb.py}. While it's probably foolish to implement one's own
database layer, the {\sf flatdb} system is simple, fairly efficient,
and was not a major sink of programmer time. The motivation for
creating it was to take advantage of the fact that the Git repository
would effectively provide change tracking and data backup. While a few
existing text-based Python database modules were found, they were
generally not well-engineered.

Additional utilities were created to populate the observation
databases more fully as certain new needs came to light. For instance,
{\sf pdb-filldatasizes} computes and records the total data size of
each session, to allow reporting of the total survey data volume. {\sf
  pdb-fillfocus} determines and inserts focus setting information
because the importance of this information was not initially obvious.

Several schedule blocks were used to observe AGILITE sources outside
of the commensal observing system. Several tools were written to
retroactively populate the database with information from these
sessions so that all of the relevant information would be centralized.
Not all sessions run outside of the commensal system map onto the
schema of the observation database, but many do. The program {\sf
  misc-scantree} performs a similar task to {\sf pdb-scansess},
printing out processed scan information, but it does not rely on the
existence of metadata from {\sf obs-slave.py}. The program {\sf
  pdb-retrosess} does the job of inserting this information into the
database. The two can be linked together in a shell pipeline.

Various other utilities query the databases to analyze, {\it e.g.},
hour angle coverage of a source, total observing time, or data
quality. The most widely-used of these tools is {\sf qdb-dumpsess},
which can summarize the status of an entire observing campaign or one
of its component sessions. The simultaneous use of two correlators
complicates some queries, since separate scan records are created for
each correlator's data stream. For instance, if one na\"ively adds
up the integration time of a group of scans on a particular source,
the total will be about twice the actual integration time, because
two correlators were active. A different database schema could
certainly trade off this particular issue for other ones.

A final group of tools integrate the information in the observation
database to ARF, the ATA Reduction Framework, the system currently
used to analyze the commensal correlator observations. While a
discussion of ARF is outside the scope of this document, I'll mention
that the programs {\sf rx-create} and {\sf rx-recreate} stub out ARF
``reduction specifications'' and link ARF work directories with the
appropriate raw datasets.

\section{Recommendations}
\label{s:recos}

I conclude with some recommendations to be kept in mind when designing
and implementing future commensal observing campaigns. These are
naturally targeted toward projects similar to the one described here
and won't apply to every campaign that could be described as
``commensal.''
\begin{itemize}
\item {\bf KISS: Keep It Simple, Stupid.} Perhaps the only universal
  engineering maxim, and it's as relevant in the commensal context as
  it is everywhere else. It's almost always better to get something
  small working and build out. We certainly had many ideas for the
  campaign described in this memo that, in retrospect, would have been
  complete wastes of time to implement.
\end{itemize}
The following group of recommendations is actually more specific {\it
  large} projects than {\it commensal} projects. For a more
authoritative perspective, \citet{kgb+08} present some lessons learned
from the operations of the Sloan Digital Sky Survey.
\begin{itemize}
\item {\bf It's all about the data.} You'll be able to write new
  software, but you won't be able to take new data, so get the data
  right. The highest priority, of course, is getting whatever allows
  you to accomplish your science goals. Beyond that, the more uniform
  your data are, the easier processing will be --- so not only is it
  important to get the data right, but it makes life a lot easier to
  think about these things hard before the campaign even starts. Large
  quantities of data are extremely slow to move around, so
  production-scale processing needs to require as little copying as
  possible.
\item {\bf Get an end-to-end processing pipeline online as soon as
  possible.} You don't want to wait until after half your observing is
  done to realize that you need a new kind of calibration observation,
  or you're missing some very useful metadata, or something's wrong in
  the archives. Once a pipeline is in place, you can also start
  inserting sanity checks to discover bad data just days, not months,
  after they start coming in. Start out by stubbing out as much as
  possible (cf. KISS) and fill in the details as you can.
\item {\bf Define and monitor observing efficiency metrics.} You want
  to know if you're on track to reach your science goals, and your
  observations will almost surely be less efficient in practice than
  in theory. Choose a just few key metrics to avoid information
  overload. As with the processing pipeline, the earlier these can be
  put into regular use, the better.
\item {\bf Schedule a time to step back and review.} If observations
  are running near-continuously, it becomes difficult to take the time
  to review how the campaign has progressed and ponder how efficiency
  might be improved. After a large project has gotten rolling, it's
  probably worthwhile to fall behind on data processing in order to
  spend a week or so conducting such a review.
\end{itemize}
These recommendations are more specific to commensal campaigns:
\begin{itemize}
\item {\bf ``Electronics before concrete.''} This is a slogan
  promoting cost-effective design adopted by Swiss railroad planners
  \citep{s08}. The idea is that it's much cheaper and faster to
  retrofit existing systems (new signaling systems on existing lines,
  in the Swiss context) than it is to build new ones from
  scratch. This is certainly also true when building the
  infrastructure to run a commensal campaign: you should take
  advantage of existing infrastructure for non-commensal
  observations. It only took a few weeks to bolt a small telemetry
  module onto the existing SETI Prelude system; meanwhile, it took
  several years to get SonATA, the from-scratch Prelude replacement,
  on the air.
\item {\bf ``Organization before electronics before concrete.''} This
  variation is used by some German planners \citep{bks11}. Their point
  is large projects often involve multiple unaffiliated actors (rail
  transit agencies) whose turf wars and internal preferences can lead
  to plans that are {\it much} costlier than what would be arrived at
  by an apolitical efficiency-focused team; thus, when planning a
  project, one of the most useful things you can do is constructively
  tackle points of conflict, even though it's always tempting to avoid
  them. In a commensal campaign, of course, there are also multiple
  actors with diverging interests. It was possible to design the
  telemetry system (\S\ref{s:synch}) in such a reliable way only
  because decisions about observing strategy were negotiated in person
  and not left to be implicitly made in software.
\item {\bf Share as much infrastructure as possible.} This is related
  to the previous item. In a commensal context, multiple systems doing
  the same job will find a way to get out of sync and cause
  problems. The scheduling system of this memo is an example of
  this. It isn't bad considered on its own, but its interactions with
  the SETI and HCRO systems are problematic. Much time has been spent
  emailing back and forth to negotiate and confirm schedule
  changes. There's no good reason for the correlator and beamformer
  observations to be scheduled separately. The {\it bad} reason for
  this is that the two sets of observations are run on separate
  computer systems and so it is difficult for them to share
  information. I have no doubt, however, that some arrangement could
  have been arrived at, avoiding not only the tedious emailing but
  also making observing more robust thanks to the elimination of {\it
    an entire class} of potential failures.
\item {\bf Make standalone pilot observations, and expect to perform
  standalone ``patch-up'' observations.} It will help the campaign
  design process if you've performed a few test observations without
  the commensal infrastructure to check out the mechanics and
  hopefully learn about any potential pitfalls. There will almost
  definitely be {\it some} kind of observation that would be good to
  get that won't be made (or will be botched) during regular
  operations, so plan to make occasional standalone observations to
  cover these gaps. (Don't forget to propose for the time to make
  these observations, if necessary!)
\item {\bf Be considerate of observatory staff.} This should go
  without saying. In the particular case of commensal campaigns, hacks
  to the observatory infrastructure will likely be necessary, and it's
  vital that these occur in a fashion acceptable and comprehensible to
  the staff who will have to deal with them.
\end{itemize}
Finally, these recommendations may be relevant to the software
implementation of a commensal campaign:
\begin{itemize}
\item {\bf Avoid bidirectional communications.} There are many more
  ways to implement a one-way communications channel than a two-way
  channel. The one-way telemetry stream described in this memo
  (\S\ref{s:synch}) was straightforward to implement and robust, as
  demonstrated by the smooth replacement of Prelude with SonATA and
  the easy creation of simple tools such as {\sf obs-master.py} or
  {\sf obs-listener.py}. The success of the synchronization protocol
  can be contrasted with that of JSDA, the messaging system used to
  interlink the ATA subsystems \citep{gkkm+10}. JSDA uses a two-way
  ``remote procedure call'' model for communications. While it
  undoubtedly offers important functionality, the JSDA system is
  complicated, and obscure lockups or communications failures have not
  been uncommon. The heaviness of the protocol also makes it
  time-consuming to integrate new software into the messaging network.
\item {\bf Use stateless protocols.} Software and hardware will
  inevitably fail, and testing of your observing systems is likely to
  be inexhaustive. Using a stateless architecture makes it so you
  don't need to even try to handle a variety of tricky problems.
\end{itemize}

Speaking as the person who drove the commensal correlator system and
who has to reduce all of the data it generated, I feel that it
performed well, and that this was largely due to some good design
decisions early on in the commensal campaign. It's not likely that
much of the particular implementation will be portable to future
facilities, but I hope that some of these recommendations will be.

\section*{Acknowledgments}

Special thanks to Peter Backus, Tom Kilsdonk, and Jon Richards for
their work to make the commensal observing campaign happen; as the
ones in charge of moving the dishes, they had a much more difficult
job than I did. Thanks too to Rick Forster, Samantha Blair, Karto
Keating, and Colby Guti\'errez-Kraybill for keeping the array running
24/7, or as close to that as was humanly possible. Thanks to Geoff
Bower for carefully reading drafts of this piece. Finally, the MMM
group was an invaluable resource for discussing all things
observational.

\bibliographystyle{yahapjhack}
\bibliography{pkgw,extra}{}

\begin{thebibliography}{7}
\expandafter\ifx\csname natexlab\endcsname\relax\def\natexlab#1{#1}\fi

\bibitem[{{Baumgartner} {et~al.}(2011){Baumgartner}, {Kantke}, \&
  {Dietz}}]{bks11}
{Baumgartner}, S., {Kantke}, T., \& {Dietz}, U.~S. 2011, Bahnknoten M\"unchen,
  \href{http://www.stadtkreation.de/munich/bahnknoten\_muenchen.html}
{{\ttfamily
  http://www.stadtkreation.de/munich/bahnknoten\_muenchen.html}}\relax

\bibitem[{Gutierrez-Kraybill {et~al.}(2010)Gutierrez-Kraybill, Keating,
  MacMahon, Williams, Harp, Ackermann, Kilsdonk, Richards, \& Barott}]{gkkm+10}
Gutierrez-Kraybill, C., Keating, G.~K., MacMahon, D., {et~al.} 2010,
  \href{http://dx.doi.org/10.1117/12.857860}{Proceedings of SPIE, 7740, 77400Z}

\bibitem[{Hyman {et~al.}(2005)Hyman, Lazio, Kassim, Ray, Markwardt, \&
  Yusef-Zadeh}]{hlk+05}
Hyman, S.~D., Lazio, T. J.~W., Kassim, N.~E., {et~al.} 2005,
  \href{http://dx.doi.org/10.1038/nature03400}{Nature, 434, 50}

\bibitem[{Kleinman {et~al.}(2008)Kleinman, Gunn, Boroski, Long, Snedden, Nitta,
  Krzesi\'{n}ski, Harvanek, Neilsen, Gillespie, Barentine, Uomoto, Tucker,
  York, \& Jester}]{kgb+08}
Kleinman, S.~J., Gunn, J.~E., Boroski, B., {et~al.} 2008,
  \href{http://dx.doi.org/10.1117/12.789612}{Proceedings of SPIE, 7016, 70160B}

\bibitem[{{Schwager}(2008)}]{s08}
{Schwager}, M. 2008, Schweizer Organisation des Bahnnetzes als Vorbild,
  \href{http://www.scritti.de/text/bahn2000.html}
{{\ttfamily http://www.scritti.de/text/bahn2000.html}}\relax

\bibitem[{{Welch} {et~al.}(2009){Welch}, {Backer}, {Blitz}, {Bock}, {Bower},
  {Cheng}, {Croft}, {Dexter}, {Engargiola}, {Fields}, {Forster},
  {Gutierrez-Kraybill}, {Heiles}, {Helfer}, {Jorgensen}, {Keating}, {Lugten},
  {MacMahon}, {Milgrome}, {Thornton}, {Urry}, {Leeuwen}, {Werthimer},
  {Williams}, {Wright}, {Tarter}, {Ackermann}, {Atkinson}, {Backus}, {Barott},
  {Bradford}, {Davis}, {Deboer}, {Dreher}, {Harp}, {Jordan}, {Kilsdonk},
  {Pierson}, {Randall}, {Ross}, {Shostak}, {Fleming}, {Cork}, {Vitouchkine},
  {Wadefalk}, \& {Weinreb}}]{theata}
{Welch}, J., {Backer}, D., {Blitz}, L., {et~al.} 2009,
  \href{http://dx.doi.org/10.1109/JPROC.2009.2017103}{IEEE Proceedings, 97,
  1438}

\bibitem[{Williams \& Bower(2010)}]{wb10}
Williams, P. K.~G., \& Bower, G.~C. 2010,
  \href{http://dx.doi.org/10.1088/0004-637X/710/2/1462}{Astrophys. J., 710,
  1462}

\end{thebibliography}

\appendix

\section{Commensal Command Summaries}
\label{s:cmds}

\begin{tabular}{rp{3.25in}}
Name & Description \cr
\hline
{\sf commensal-mc} & Monitor \& control of commensal obs by {\sf obs} user \cr
{\sf commensal-mc-helper} & Helper for the above \cr
{\sf misc-archdir} & Print the {\sf /ataarchive} directory for a
  session's data \cr
{\sf misc-crsync} & Copy session data from {\sf strato} to {\sf cosmic} \cr
{\sf misc-decodeunixtime} & Print a Unix time in human-friendly format \cr
{\sf misc-diffunixtime} & Print the difference between two Unix times
  in a human-friendly format \cr
{\sf misc-dsetnames} & Print the names of datasets including a
  particular source  \cr
{\sf misc-latestdircmd} & Print an {\sf eval}'able command to change to
  the directory containing the latest observations \cr
{\sf misc-makeunixtime} & Convert calendar information into a Unix time \cr
{\sf misc-parseplog} & Summarize an ongoing commensal observation from its
  pointing logfile \cr
{\sf misc-scantree} & Scan an arbitrary tree of visibility data \cr
{\sf misc-sessalias} & Given a session UID, print its alias \cr
{\sf misc-sessuid} & Given a session alias, print its UID \cr
{\sf misc-whatsnear} & Given coordinates, find the nearest observed pointings \cr
{\sf obs-hourly} & The hourly observing launcher \cr
{\sf obs-kickstart} & Launch observations right now \cr
{\sf obs-launcher} & Backend program to launch observations \cr
{\sf obs-launcher-helper} & Helper for the above \cr
{\sf obs-listener.py} & Debug the commensal synchronization protocol \cr
{\sf obs-master.py} & Drive the array simplistically \cr
{\sf obs-slave.py} & Perform commensal correlator observations \cr
{\sf pdb-datealias} & Set session aliases from their observing dates \cr
{\sf pdb-filldatasizes} & Fill in data size information for sessions
  missing it \cr
{\sf pdb-fillfocus} & Fill in focus information for sessions missing it \cr
{\sf pdb-filllsts} & Fill in LST information for sessions missing it
\end{tabular}


\begin{tabular}{rp{3.25in}}
Name & Description \cr
\hline
{\sf pdb-fillobscfgs} & Fill in observing frequency/focus configuration
  information for sessions missing it \cr
{\sf pdb-importagcts} & Import database information from the first-generation
  AGCTS database \cr
{\sf pdb-importproj} & Import information from one commensal observing
  project (campaign) into another \cr
{\sf pdb-retrosess} & Retroactively import the scans and metadata for a
  session \cr
{\sf pdb-scansess} & Import scan information for a given session \cr
{\sf pdb-stubsess} & Stub out session information in the database from
  the current schedule \cr
{\sf plan-archdir} & Print out the expected archive directory of a scheduled
  observing session \cr
{\sf plan-importtxt} & Import the ATA textual schedule into the commensal
  schedule \cr
{\sf plan-schedsess} & Manually schedule an observing block \cr
{\sf plan-showsched} & Print the current observing schedule \cr
{\sf qdb-dumpsess} & Print information about an observing campaign or
  session \cr
{\sf qdb-sesssrchacov} & Show how well each session covers a given source
  in a given hour angle range \cr
{\sf qdb-srccov} & Summarize the coverage of a particular source in a
  project on a session-by-session basis \cr
{\sf qdb-srchacov} & Show the hour angle coverage of a given source over
  the course of the campaign \cr
{\sf qdb-summarize} & Summarize the overall observing statistics of a
  campaign \cr
{\sf qdb-toscan} & Print out a list of sessions that probably need to
  be scanned \cr
{\sf rx-create} & Create an ARF reduction specification and workspace
  for a given session \cr
{\sf rx-recreate} & Re-realize an ARF reduction workspace for a given
  session\cr
{\sf rx-status} & Set the reduction status of a session \cr
{\sf rx-suggest} & Suggest a useful session to reduce
\end{tabular}

\section{Commensal Python Module Summaries}
\label{s:modules}

\begin{tabular}{rp{3.5in}}
Name & Description \cr
\hline
{\sf catalog} & Loading and using source catalogs \cr
{\sf files} & Computing paths for accessing file resources \cr
{\sf flatdb} & Simple line-oriented textual database  \cr
{\sf projdb} & Session and scan database implementation \cr
{\sf protocol} & Synchronization protocol implementation \cr
{\sf sched} & Loading and using observing schedules \cr
{\sf util} & Miscellaneous utilities
\end{tabular}

\end{document}